# Inference Scaling Reshapes AI Governance

Toby Ord[*]

Oxford Martin AI Governance Initiative
University of Oxford

The shift from scaling up the pre-training compute of AI systems to scaling up their inference compute may have profound effects on AI governance. The nature of these effects depends crucially on whether this new inference compute will primarily be used during external deployment or as part of a more complex training programme within the lab. Rapid scaling of inference-at-deployment would: lower the importance of open-weight models (and of securing the weights of closed models), reduce the impact of the first human-level models, change the business model for frontier AI, reduce the need for power-intense data centres, and derail the current paradigm of AI governance via training compute thresholds. Rapid scaling of inference-during-training would have more ambiguous effects that range from a revitalisation of pre-training scaling to a form of recursive self-improvement via iterated distillation and amplification.

## The end of an era — for both training and governance

The intense year-on-year scaling up of AI training runs has been one of the most dramatic and stable markers of the Large Language Model era. Indeed it had been widely taken to be a permanent fixture of the AI landscape and the basis of many approaches to AI governance.

But recent reports from unnamed employees at the leading labs suggest that their attempts to scale up pre-training substantially beyond the size of GPT-4 have led to only modest gains which are insufficient to justify continuing such scaling and perhaps even insufficient to warrant public deployment of those models (Hu & Tong, 2024). A possible reason is that they are running out of high-quality training data. While the scaling laws might still be operating (given sufficient compute and data, the models would keep improving), the ability to harness them through rapid scaling of pre-training may not. What was taken to be a fixture may instead have been just one important era in the history of AI development; an era which is now coming to a close.

Just before the reports of difficulties scaling pre-training, OpenAI announced their breakthrough reasoning model, o1 (OpenAI 2024). Their announcement came with a chart showing how its performance on a difficult mathematics benchmark could be increased via scaling compute dedicated to post-training reinforcement learning (to improve the overall performance of the model); or by scaling the inference compute used on the current task.

---

[*] Special thanks to Seb Krier for discussions that inspired some of these ideas.

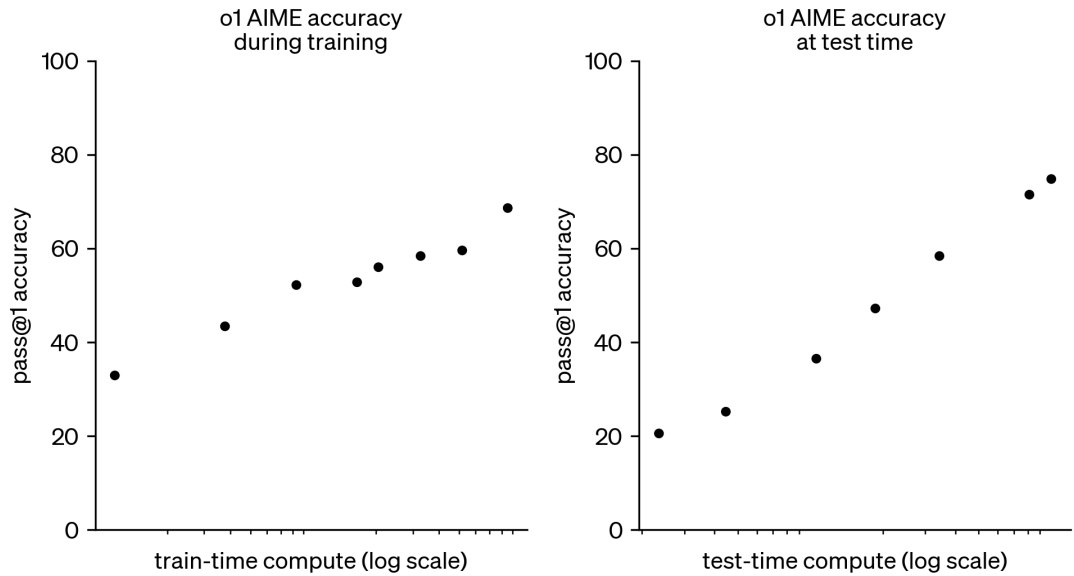

*Figure 1.* How o1's performance on AIME scales with post-training compute and with inference compute, from OpenAI (2024).

This has led to intense speculation that the previous era of scaling pre-training compute could be followed by an era of scaling up inference-compute. In this essay, I explore the implications of this possibility for AI governance.

In some ways a move to scaling of inference compute may be a continuation of the previous paradigm (as lab leaders have been suggesting†). For example, work on the trade-off between pre-training compute and inference compute suggests that (on the current margins) increasing inference compute on the task at hand by 1 order of magnitude often improves performance as much as increasing pre-training compute by 0.5 to 1 orders of magnitude (Jones 2021, Villalobos & Atkinson 2023). So we may be tempted to see it simply as an implementation detail in the bigger story of scaling up compute in general.

---

† e.g. Dario Amodei has said 'Every once in a while, the underlying thing that is being scaled changes a bit, or a new type of scaling is added to the training process. From 2020-2023, the main thing being scaled was *pretrained models*: models trained on increasing amounts of internet text with a tiny bit of other training on top. In 2024, the idea of using *reinforcement learning* (RL) to train models to generate chains of thought has become a new focus of scaling.'

But a closer look suggests that may be a mistake. There are a number of key differences between scaling pre-training and scaling inference — both for the labs and for AI governance.

I shall argue that many ideas in AI governance will need either an adjustment or an overhaul. Those of us in the field need to look back at the long list of ideas we work with and see how this affects each one.

There is a lot of uncertainty about what is changing and what will come next.

One question is the rate at which pre-training will continue to scale. It may be that pre-training has topped out at a GPT-4 scale model, or it may continue increasing, but at a slower rate than before. Epoch AI (2024) suggests the compute used in LLM pre-training has been growing at about 5x per year from 2020 to 2024. It seems like that rate has now fallen, but it is not yet clear if it has gone to zero (with AI progress coming from things other than pre-training compute) or to some fraction of its previous rate.

A second — and ultimately more important — question concerns the nature of inference-scaling. We can view the current AI pipeline as pre-training (such as via next-token prediction), followed by post-training (such as RLHF or RLAIF), followed by deploying the trained model on a vast number of different tasks (such as through a chat interface or API calls).

The second question is whether the scaled-up inference compute will primarily be spent during deployment (like in o1 and R1) or as part of a larger and more complex post-training process (like the suggestions that OpenAI may have used trained o3 via many runs of o1). Each of these possibilities has important — but different — implications for AI governance.

## Scaling inference-at-deployment

Let's first consider the scenario where for the coming years, the lion's share of compute scaling goes into scaling up the inference compute used at deployment. In this scenario, the pre-trained system is either stuck at GPT-4 level or only slowly progressing beyond that, while new capabilities are being rapidly unlocked via more and more inference compute. Some of this may be being spent in post-training as the system learns how to productively reason for longer times (e.g. the reinforcement learning in the left-hand chart of OpenAI's o1 announcement), but for this scenario, we are supposing that this one-off cost is comparatively small and that the main thing being scaled is the deployment compute.

In this scenario, we may be able to use rules of thumb such as

Effective orders of magnitude = OOMs of pre-training + 0.7 × OOMs of inference to estimate the capabilities of an inference-scaled model in terms of the familiar yardstick of pure pre-training. But overreliance on such formulas could obscure key changes in the new scaling paradigm — changes that stem from the way the benefits of inference-at-deployment depend upon the task at hand, the way the amount of inference can be tuned to the task, and the way the costs shift from training time to deployment time.

**Reducing the number of simultaneously served copies of each new model**

It currently takes a vast number of chips to train a frontier model. Once the model is trained, those chips can be used for inference to deploy a large number of simultaneous copies of that model. Dario Amodei (2024) of Anthropic estimates this to be 'millions' of copies. This number of copies is a key parameter for AI governance as it affects the size of the immediate impact on the world the day the new model is ready. A shift to scaling inference-at-deployment would lower this number. e.g. if inference-at-deployment is scaled by two orders of magnitude, then this key parameter goes down by a factor of 100, and the new model can only be immediately deployed in 1% as many tasks as it would be if it had been scaled by pre-training compute.‡

**Increasing the price of first human-level AGI systems**

A related parameter is how expensive the first 'human-level' AI systems will be to run. By previous scaling trends we might expect the first such systems to cost much less than human labour, meaning that they could be immediately deployed at a great profit, which could be ploughed back into renting chips to run more copies of them in an escalating feedback loop. But each additional order of magnitude that goes to inference-at-deployment may increase the cost of using these systems by up to an order of magnitude.

This will blunt the immediate impact of reaching this threshold and may even be enough such that there is an initial period where we first have access to 'human-level' AGI systems at more than the cost of equivalent human labour. If so, such systems could be available to study (for safety work) or demonstrate (before the world's leaders) before they have transformative effects on society.

Obviously the fact that AI is already much better than humans at some tasks while much worse at others complicates this idea of reaching 'human-level', but I believe it is still a useful lens. For example, you can ask whether the first systems that can perform a particular job better than humans will cost more or less than human wages for that job.

---

‡ Or, somewhat equivalently, it might be better thought of as slowing these systems down by that factor (e.g. 100x). Amodei's estimate is that AI systems are currently 10x–100x human speed, but if they reach intelligence via inference scaling, they may be slower than humans. Both ways of looking at it lead to the same reduction in the 'human-days-equivalent of AI work each day' when the systems are switched from training to deployment.

**Reducing the value of securing model weights**

Suppose that for frontier models, training compute plateaus at something like the GPT-4 level while inference-at-deployment scales by a factor of 100. Then the value of stealing model weights hasn't increased over time — it is just the value of not having to train a GPT-4 level model (which has been decreasing over time by about 4x per year due to algorithmic efficiency improvements and Moore's law (Epoch AI 2024)). And even if the weights were stolen, the thief would still have to pay the high inference-at-deployment costs. If they intend to use the model at anything like the scale current leading models are used, these would be the lion's share of the total costs and much higher than the training costs of the model they stole.

**Reducing the benefits and risks of open-weight models**

This also affects both the benefits and drawbacks of open-weight models. If open-weight models require vast amounts of inference-at-deployment from their users, then they are much less attractive to those users than models of equivalent capability that were entirely pre-trained (since then the model trainer has paid those costs for you). So open-weight models could become much less valuable for their users and also less dangerous in terms of proliferation of dangerous capabilities. They would become less strategically important overall.

**Unequal performance for different tasks and for different users**

Scaling inference-at-deployment helps with some kinds of tasks much more than others. It helps most with tasks where the solution is objectively verifiable, such as certain kinds of maths and programming tasks. It can also be useful for tasks involving many steps. Two good heuristics for the tasks that benefit from inference scaling are:

1. tasks that benefit from System 2–type thinking (methodical reasoning) when performed by humans,

2. tasks that typically take humans a long time (as this shows these tasks can benefit from a lot of thinking before diminishing marginal returns kick in).

Because some tasks benefit more from additional inference than others, it is possible to tailor the amount of inference compute to the task, spending 1,000x the normal amount for a hard, deep maths problem, while just spending 1x on problems that are more intuitive. This kind of tailoring isn't possible with pre-training scaling, where scaling up by 10x increases the costs for everything.

A related change is that users with more money will likely be able to convert that into better answers. We've already seen this start to happen at OpenAI (the first frontier company to allow access to a model that scales inference-at-deployment). They now charge 10x as much for access to the version using the most inference-compute. We'd become accustomed to a dollar a day getting everyone the same quality of AI assistance. It was as Andy Warhol said about Coca Cola:

> 'What's great about this country is that America started the tradition where the richest consumers buy essentially the same things as the poorest. … the President drinks Coca Cola, Liz Taylor drinks Coca Cola, and just think, you can drink Coca Cola, too. A coke is a coke and no amount of money can get you a better coke than the one the bum on the corner is drinking.'

But scaling inference-at-deployment ends that.

**Changing the business model and industry structure**

The LLM business model has had a lot in common with software: big upfront development costs and then comparatively low marginal costs per additional customer. Having a marginal cost per extra user that is lower than the average cost per user encourages economies of scale where each company is incentivised to set low prices to acquire a *lot* of customers, which in turn tends to create an industry with only a handful of players. But if the next two orders of magnitude of compute scale-up go into inference-at-deployment instead of into pre-training, then this would change, upsetting the existing business model and perhaps allowing more room for smaller players in the industry.

**Reducing the need for monolithic data centres**

While training compute benefits greatly from being localised in the same data centre, inference-at-deployment can be much more easily spread between different locations. Thus if inference-at-deployment is being scaled by several orders of magnitude, it could avoid current bottlenecks concerning single large data centres, such as the need for a large amount of electrical power draw in a single place (which has started to require its own large powerplant). So if one hoped for the government to be able to exert some control over AI labs via the carrot of accelerated power plant approvals, inference-at-deployment may change that. And it will make it harder for governments to keep track of all the frontier models being trained by tracking the largest datacentres.

**Breaking the strategy of AI governance via compute thresholds**

Many AI governance frameworks are based around regulating only those models above a certain threshold of training compute. For example, the EU AI Act uses $10^{25}$ FLOP while the US executive order uses $10^{26}$ FLOP. This allows them to draw a line around the few potentially dangerous systems without needing to regulate the great majority of AI models (Heim & Koessler, 2024). But if capabilities can be increased via scaling inference-at-deployment then a model whose training compute was below these thresholds might be amplified to become as powerful as those above them. For example, a model trained with $10^{24}$ FLOP might have its inference scaled up by 4 OOM and perform at the level of a model trained with $10^{27}$ FLOP. This threatens to break this entire approach of training-compute thresholds.

At first the threat might be that someone scales up inference-at-deployment by a very large factor for a small number of important tasks. If the inference scale-up is only happening on a small fraction of all tasks the model is deployed on, one could use a very high scale-up factor (such as 100,000x) and suddenly operate at the level of a new tier of model.

The main limitation on this at the moment is that many current techniques for inference scaling seem to hit plateaus that can't be exceeded by any level of inference scale-up. Exceeding these plateaus requires substantial research time by AI scientists and engineers, such that if someone tried to use a GPT-4 level model with 100,000x the inference compute, it may not be able to make good use of most of that compute. However, labs are developing better ways to use large multipliers of inference compute before reaching performance plateaus and this work is proceeding very quickly. For example, OpenAI demonstrated their o3 model making use of 10,000x as much compute as their smallest reasoning model, o1-mini, and so presumably an even larger factor above their base model, GPT-4o.

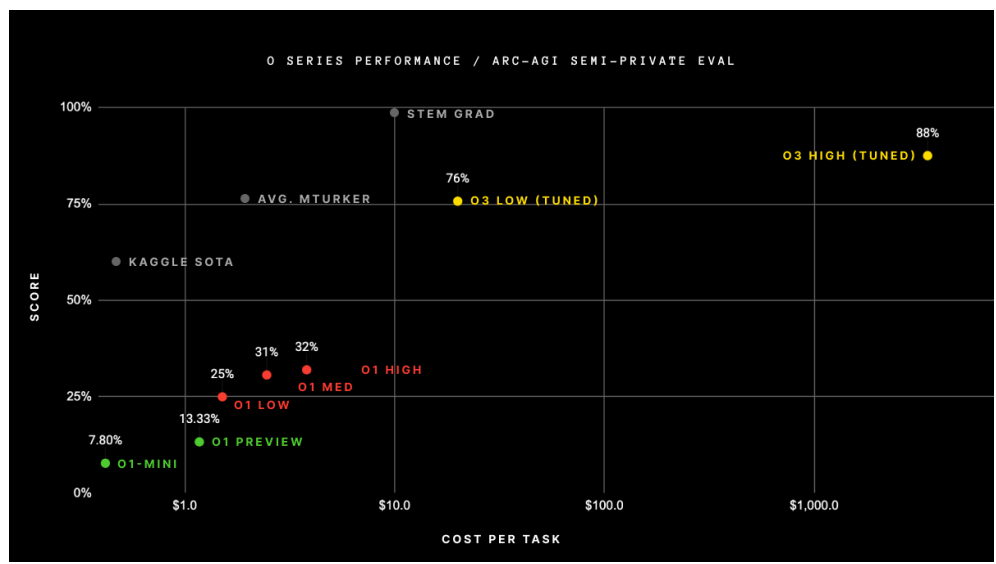

*Figure 2.* Scaling up compute by a factor of 10,000 without hitting a plateau, from arcprize.org.

Leading labs have also been scaling their data centres and improving algorithmic efficiency such that they may already have 100x the effective-compute of the first data centres capable of serving GPT-4 to customers. This would allow more than just a few people to use versions with greatly scaled-up inference-at-deployment. For example, OpenAI's recently launched deep research model (based on o3) may well exceed the performance of a system pre-trained on $10^{26}$ FLOP, even if it is technically below that threshold.

While one could try to change the governance threshold to incorporate the inference-at-deployment as well as the pre-training compute, this would face serious problems. The current framework aims to separate AI systems that could be dangerous from those that can't be. It aims to regulate dangerous objects, not dangerous uses of objects. But a revised threshold would depend not just on the model but on how you are using it, which would be a different and more challenging kind of governance threshold.

Perhaps one way to save the compute thresholds is to say that they cover both systems above $10^{26}$ FLOP of pre-training and systems above some smaller threshold (e.g. $10^{24}$ FLOP of pre-training) that have had post-training to allow themselves to benefit from high inference-at-deployment. But this still suffers from increased complexity and fewer bright lines.

Overall, scaled up inference-at-deployment looks like a big challenge for governance via compute thresholds.

**Scaling inference-during-training**

AI labs may also be able to reap tremendous benefit from these inference-scaled models by using them as part of the training process. If so, the large scale-up of compute resources could go into post-training rather than deployment. This would have very different implications for AI governance.

In this section, we'll focus on the implications of a pure strategy of using inference-scaling *only* during the training process. This will clarify what it contributes to the overall picture of AI governance, though realistically we will see inference-scaling in both training and deployment.

An obvious approach to scaling inference-during-training is to use an inference-scaled model to generate large amounts of high-quality synthetic data on which to pre-train a new base model. This would make sense if the challenges in scaling up pre-training beyond GPT-4 are due to running out of high-quality training data. For example, court documents (Kadrey v. Meta Platforms, 2023) have revealed that Meta's Llama3 team decided to train on an illegal Russian repository of copyrighted books, LibGen, because they were unable to reach GPT-4 level without it.:

> 'Libgen is essential to meet SOTA [state-of-the-art] numbers, across all categories, and it is known that OpenAI and Mistral are using the library for their models (through word of mouth).'

This strongly suggests that even though there are still many more unused tokens on the indexed web (about 30x as many as are used in GPT-4 level pre-training), performance is being limited by lack of high-quality tokens. There have already been attempts to supplement the training data with synthetic data (data produced by an LLM), but if the issue is more about quality than raw quantity, then they need the best synthetic data they can get.

Inference-scaling can help with this by boosting the capability of the model producing the synthetic data. One way to do this is via domains such as mathematics or programming where one can tell whether a generated solution is correct and how efficient it is. The training programme could involve generating lots of proofs and computer programs using advanced reasoning models until it finds high quality solutions, and then add those to the stock of data that goes into pre-training the next base model.

This access to ground truth in mathematical disciplines is particularly important for getting the right training signal. But even for domains that are less black and white, it may be possible to trade extra inference compute for better synthetic data. For example, one could generate many essays, run several rounds of editing on them, and then assess them for originality, importance of insight, and lack of detectable errors, putting only those of the highest quality into the stock of synthetic data.

Relatedly, one could apply this technique to the stock of human-generated training data, assessing each document in the training data and discarding those that are below-average in quality. This could either improve the average quality of the data they already use or make some fraction of the unused sources of data usable.

On its own, this approach of scaling inference-during-training to produce synthetic data for pre-training is not so interesting from an AI governance perspective. Its main direct effect is to allow the scaling of pre-training compute to recommence, breathing new life into the existing scaling paradigm.

But there is a modification of this approach that may have the potential to lead to explosive growth in capabilities. The idea is to rapdily improve a model's abilities by amplifying its abilities (through inference scaling) then distilling those into a new model, and repeating this process many times. This idea is what powered the advanced self-play in DeepMind's AlphaGo Zero (Silver et al. 2017), and was also independently discovered by Anthony et al. (2017) and, in the context of AI-safety, by Christiano (2017).

In the case of AlphaGo Zero, you start with a base model, $M_0$, that takes a representation of the Go board and produces two outputs: a probability distribution over the available moves (representing the chance a skilled player would choose them) and a probability representing the chance the active player will eventually win the game.§ This model will act as an intuitive 'System 1' approach to game playing, with no explicit search.

The training technique then plays 25,000 games of Go between two copies of $M_0$ amplified by a probabilistic technique for searching through the tree of moves and countermoves called Monte Carlo Tree Search (MCTS). That is, both players use MCTS with $M_0$ guiding the search by using its estimates of likely moves and position strength as a prior. By repeatedly calling $M_0$ in the search (thousands of times), we get a form of inference-scaling which amplifies the power of this model. We could think of it as taking the raw System 1 intuitions of the base model and embedding them in a System 2 reasoning process which thinks many moves ahead.

This amplified model is better than the base model at predicting the move most likely to win in each situation, but it is also much more costly. So we train a new model, $M_1$, to predict the outputs of $M_0$ + search. Following Christiano, I shall call this step distillation, though in the case of AlphaGo Zero, $M_1$ was simply $M_0$ with an additional stage of training. This trained its move predictions to be closer to the probability distribution over moves that $M_0$ + search gives and trained its board evaluations to be closer to the final outcome of the self-played games. While $M_1$ won't be quite as good at Go as the amplified version of $M_0$, it is better than $M_0$ alone.

But why stop there? We can repeat this process, amplifying $M_1$ through inference-scaling by using *it* to guide the search process, producing a level of play beyond any seen so far ($M_1$ + search). This then gets distilled into a new model, $M_2$, and we proceed onwards and upwards, climbing higher and higher along the ladder of Go-playing performance.

---

§ For AlphaGo Zero, the goal was to start with zero information about Go and learn everything, so $M_0$ was simply a randomly initialised network. But it is also possible to start with a more advanced network as $M_0$, such as one trained to imitate human behaviour.

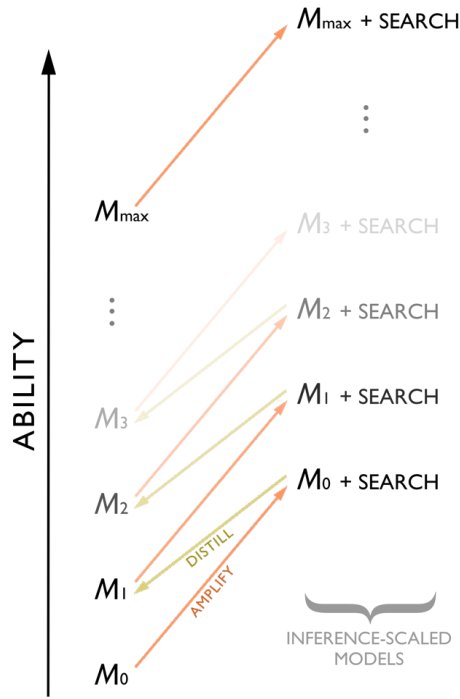

*Figure 3.* AlphaGo Zero's training via alternating amplification with inference-scaling and distillation of this knowledge back into the main network.

After just 36 hours the best model (with search) had exceeded the ability of AlphaGo Lee (the version that beat world-champion Lee Sedol) which had been trained for *months* but lacked some innovations including this structure of iterated amplification and distillation. Within 72 hours AlphaGo Zero was able to defeat AlphaGo Lee by 100 games to zero. And after 40 days of training (and 4.9 million games of self-play[**]) it reached its performance plateau, $M_{max}$, where the unamplified model could no longer meaningfully improve its predictions of the amplified model. At this point the amplified version had an estimated Elo rating of 5,185 — far beyond the 3,739 of AlphaGo Lee or the low 3,000s of the world's best human players. Even when the final model was used without any search process (i.e. without any scaling of inference-at-deployment), it achieved a rating of 3,055 — roughly at the level of a human pro player despite playing from pure 'intuition' with no explicit reasoning.

It may be possible to use such a process of iterated distillation and amplification in the training of LLMs. The idea would be to take a model such as GPT-4o (which applied a vast amount of pre-training to provide it with a powerful System 1) and use it as the starting model, $M_0$.[††] Then amplify it via inference scaling into a model that uses a vast number of calls to $M_0$ to simulate System 2–type internal reasoning before returning its final answer (as o1 and R1 do). Then distill this amplified model into a new model, $M_1$, that is better able to produce the final answer from the amplified model without doing the hidden reasoning steps.‡‡ If this works, you now have model that is more capable than GPT-4o without using extra inference-at-deployment.

---

[**] Given 4.9 million games of self-play and a set-up with 25,000 games before each distillation, there were presumably 1,960 iterations of amplification and distillation before it reached its plateau, such that $M_{max}$ is $M_{1,960}$.

[††] Like o1 and R1, we would presumably include additional RL post-training to prepare it for use in inference scaling.

By iterating this process of amplification followed by distillation, it may be possible for the LLM (like AlphaGo Zero) to climb a very long way up this ladder before the process runs out of steam. And the time for each iteration may be substantially shorter than the time between major new pre-training runs. Like AlphaGo Zero, the final distilled model could display very advanced capabilities even without amplification. If this all worked, it would be a way of scaling inference-during-training to substantially quicken the rate of AI progress.

It is not at all clear whether this would work. It may plateau quickly, or require rapidly growing parameter counts to distill each new model, or take too long per step, or too many steps, or require years worth of engineering effort to overcome the inevitable obstacles that arise during the process.§§ But AlphaGo Zero does give us a proof of concept of a small team at a leading lab achieving take-off with such a process and riding its rapid ascent to reach capabilities far beyond the former state-of-the-art.

So iterated distillation and amplification provides a plausible pathway for scaling inference-during-training to rapidly create much more powerful AI systems. Arguably this would constitute a form of 'recursive self-improvement' where AI systems are applied to the task of improving their own capabilities, leading to a rapid escalation. While there have been earlier examples of this, they have often been on narrow domains (e.g. the game of Go) or have only applied to certain cognitive abilities (e.g. 'learning how to learn') and so been bottlenecked on other abilities. Iterated distillation and amplification of LLMs is a version that could credibly learn to improve its own general intelligence.

What does this mean for AI governance? A key implication is that scaling inference-during-training may mean we have less transparency into the best current models. While this use of inference inside the training process would reach the EU AI Act's compute threshold, that threshold only requires oversight when the model is deployed.*** Thus it may be possible for companies to substantially scale up the intelligence of their leading models without anyone outside knowing. AI governance may then have to proceed from a state of greater uncertainty about the state of the art. Relatedly, the lack of transparency would mean the public and policymakers wouldn't be able to try these state-of-the-art models and so the overton window of available policy responses won't be able to shift in response to them. This would lead to less regulation and a more abrupt shock to the world when the models at the top of the training ladder are deployed.

---

‡‡ Here $M_1$ could be a fresh model distilled from the inference-scaled $M_0$, or it could be $M_0$ with fine-tuning to make it behave more like the inference-scaled $M_0$.

§§ It is also possible that it will work in some domains (such as mathematics and coding) but not others, leading to superhuman capabilities in several new domains, but not across the board.

But perhaps most importantly, the possibility of training general models via iterated distillation and amplification could shorten the timelines until AGI systems with transformative impacts on the world. If this was combined with a lack of transparency about state of the art models during internal scaling, we couldn't know for sure if timelines were shortening or not, making it hard to know whether emergency measures were required. All of this suggests that policies to require disclosure of current capabilities (and immediate plans for greater capabilities) would be very valuable.

**Conclusions**

The shift from scaling pre-training compute to scaling inference compute may have substantial implications for AI governance.

On the one hand, if much of the remaining scaling comes from scaling inference at deployment, this could have implications including:

- Reducing the number of simultaneously served copies of each new model
- Increasing the price of first human-level AGI systems
- Reducing the value of securing model weights
- Reducing the benefits and risks of open-weight models
- Unequal performance for different tasks and for different users
- Changing the business model and industry structure
- Reducing the need for monolithic data centres
- Breaking the strategy of AI governance via compute thresholds

On the other hand, if companies instead focus on scaling up the inference-during-training, then they may be able to use reasoning systems to create the high-quality training data needed to allow pre-training to continue. Or they may even be able to iterate this in the manner of AlphaGo Zero and scale faster than ever before — up the ladder of iterated distillation and amplification. This possibility may lead to:

- Less transparency into the state of the art models
- Less preparedness among the public and policymakers
- Shorter timelines to transformative AGI

---

*** And only when deployed inside the EU itself, where OpenAI's inference-scaled model deep research is conspicuously absent.

Either way, the shift to inference-scaling also makes the future of AI less predictable than it was during the era of pre-training scaling. Now there is more uncertainty about how quickly capabilities will improve and which longstanding features of the frontier AI landscape will still be there in the new era. This uncertainty will make planning for the next few years more difficult for the frontier labs, for their investors, and for policymakers. And it may provide a premium on agility: on the ability to first spot what is happening and pivot in response.

All of this analysis should be taken just as a starting point for the effects of inference-scaling on AI governance. As this transition continues it will be important for the field to track the which types of inference-scaling are happening and thus better understand which of these issues we are facing.

## Appendix. Comparing the costs of scaling pre-training vs inference-at-deployment

Scaling up pre-training by an order of magnitude and scaling up inference-at-deployment by an order of magnitude may have similar effects on the capabilities of a model, but they can have quite different effects on the total compute cost of the project. Which one is more expensive depends on the circumstances in a rather complex way.

Let's focus on the total amount of compute used for an AI system over its lifetime as the cost of that system (though this is not the only thing one might care about). The total amount of compute used for an AI system is equal to the amount used in training plus the amount used in deployment:

$C = C_{\text{pre-training}} + C_{\text{post-training}} + C_{\text{deployment}}$

Let $N$ be the number of parameters in the model, $D$ be the number of data tokens it is trained on, $d$ be the number of times the model is deployed (e.g. the number of questions it is asked) and *I* be the number of inference steps each time it is deployed (e.g. the number of tokens per answer). Then this approximately works out to:[†††]

$$C \approx ND + C_{\text{post-training}} + dNI$$

Note that scaling up the number of parameters, *N*, increases both pre-training compute and inference compute, because you need to use those parameters each time you run a forward pass in your model. But scaling up *D* doesn't directly affect deployment costs. Some typical rough numbers for these variables in GPT-4 level LLMs are:

$N = 10^{12}$     $D = 10^{13}$     $I = 10^{3}$     $d = ?$

On this rough arithmetic, the deployment costs overtake the pre-training costs when the total number of tokens generated in deployment (*dI*) is greater than the total number of training tokens *D*. That would require $d > 10^{10}$. Apparently, this is usually the case, with deployment compute exceeding total training compute on commercial frontier systems.[‡‡‡]

The most standard way of training LLMs to minimise training compute involves scaling up *N* and *D* by the same factor (Hoffmann et al., 2022). For example, if you scale up training compute by 1 OOM, that means 0.5 OOMs more parameters and 0.5 OOMs more data. So scaling up training compute by 1 OOM also increases deployment compute by 0.5 OOM. In contrast, scaling up inference-at-deployment by an order of magnitude doesn't (directly) affect pre-training compute.

When either the pre-training compute (*ND*) or the deployment compute (*dNI*) is the bulk of the total (including $C_{\text{post-training}}$), there are some simple approximations for the costs of scaling. If $C_{\text{pre-training}} \gg C_{\text{post-training}} + C_{\text{deployment}}$, then scaling pre-training by 10x increases costs by nearly 10x, while scaling inference-at-deployment (*I*) by 10x doesn't affect the total much. Whereas if $C_{\text{deployment}} \gg C_{\text{pre-training}} + C_{\text{post-training}}$, then scaling pre-training by 10x increases costs by ~3x (from the larger number of parameters needed at deployment), while scaling inference-at-deployment by 10x increases costs by nearly 10x. So there is some incentive to balance these numbers where possible.

It is important to note that the costs of scaling inference-at-deployment depend heavily on how much deployment you are doing. If you just use the model to answer a single question, then you could scale it all the way until it generates as many

---

[†††] I'm simplifying some of the details (such as the precise coefficients for each term) to make the overall structure of the equation clearer.

[‡‡‡] This has led to methods of training that use more training compute than is Chinchilla-optimal because the smaller model leads to compensating savings on the deployment compute

tokens as you pretrained on (i.e. trillions) before it appreciably affects your overall compute budget. While if you are scaling up the inference used for every question, your overall compute budget could be affected even by a 2x scale up.

**References**

Dario Amodei. 2024. Machines of Loving Grace: How AI Could Transform the World for the Better. Published online at darioamodei.com. Retrieved from: 'https://darioamodei.com/machines-of-loving-grace' [online resource]

Thomas Anthony, Zheng Tian, and David Barber. 2017. Thinking Fast and Slow with Deep Learning and Tree Search, arXiv:1705.08439 [cs.AI].

Paul Christiano. 2017. Benign model-free RL. Published online at ai-alignment.com. Retrieved from: 'https://ai-alignment.com/benign-model-free-rl-4aae8c97e385' [online resource]

Epoch AI. 2024. Key Trends and Figures in Machine Learning. Published online at epochai.org. Retrieved from: 'https://epochai.org/trends' [online resource]

Daya Guo, Dejian Yang, Haowei Zhang et al. 2025. DeepSeek-R1: Incentivizing Reasoning Capability in LLMs via Reinforcement Learning, arXiv:2501.12948 [cs.CL].

Lennart Heim and Leonie Koessler. 2024. Training Compute Thresholds: Features and Functions in AI Regulation. arXiv:2405.10799v2 [cs.CY].

J. Hoffmann, S. Borgeaud, A. Mensch, E. Buchatskaya, T. Cai, E. Rutherford, D. d. L. Casas, L. A. Hendricks, J. Welbl, A. Clark, et al. 2022. Training compute-optimal large language models, arXiv:2203.15556 [cs.CL].

Krystal Hu and Anna Tong. 2024. OpenAI and others seek new path to smarter AI as current methods hit limitations, *Reuters*, 15 Nov.

Andy L. Jones. 2021. Scaling Scaling Laws with Board Games. arXiv:2104.03113v2 [cs.LG].

Kadrey v. Meta Platforms, Inc. 2023. 3:23-cv-03417, (N.D. Cal.), Exhibit K to N. Vo Declaration, 14 Jan 2025.

J. Kaplan, Sam McCandlish, T. Henighan, Tom B. Brown, Benjamin Chess, R. Child, Scott Gray, Alec Radford, Jeff Wu, and Dario Amodei. 2020. Scaling Laws for Neural Language Models, arXiv:2001.08361 [cs.LG].

OpenAI, 12 Sep 2024. Learning to Reason with LLMs, Published online at openai.com. Retrieved from: https://openai.com/index/learning-to-reason-with-llms/

Silver, D., Schrittwieser, J., Simonyan, K. et al. 2017. Mastering the game of Go without human knowledge. *Nature* 550, 354–359. https://doi.org/10.1038/nature24270

Pablo Villalobos and David Atkinson. 2023. Trading Off Compute in Training and Inference. Published online at epoch.ai. Retrieved from: 'https://epoch.ai/blog/trading-off-compute-in-training-and-inference' [online resource].